\documentstyle[epsfig,12pt]{article}

\def\oneqt{{\textstyle{\frac{1}{4}}}}

\newcommand{\be}{\begin{equation}}
\newcommand{\ee}{\end{equation}}
\newcommand{\ba}{\begin{eqnarray}}
\newcommand{\ea}{\end{eqnarray}}
\newcommand{\bd}{\begin{displaymath}}
\newcommand{\ed}{\end{displaymath}}

\title{{\bf Finite Temperature Solitons in Non-Local Field Theories from
p-Adic Strings}}

\author{Tirthabir Biswas$^1$, Jose A. R. Cembranos$^{2,3}$ Joseph I. Kapusta$^3$}

\begin{document}

\maketitle

\begin{center}
{\it $^1$Department of Physics,
St. Cloud State University, St. Cloud, MN 56301}\\
{\it $^2$William I. Fine Theoretical Physics Institute,
University of Minnesota, Minneapolis, MN 55455}\\
{\it $^3$School of Physics and Astronomy,
University of Minnesota, Minneapolis, MN 55455}

\end{center}

\vspace{1cm}

\begin{abstract}
Non-local field theories which arise from $p$-adic string theories have vacuum soliton solutions.  We find the soliton solutions at finite temperature.  These solutions become important for the partition function when the temperature exceeds $m_s/g_o^2$ where $m_s$ is the string mass scale and $g_o$ is the open string coupling. 
\end{abstract}

\vspace{0.5cm} 

\newpage

\section{Introduction}

There is a special class of non-local field theories characterized by an infinite number of derivative terms in the form of an exponential.  These field theories describe the open string tachyon in $p$-adic string theories \cite{olson}-\cite{frampton}.  In several recent papers we studied these non-local theories at finite temperature \cite{us1,us2}.  These theories have no true particle degrees of freedom.  All contributions to the equation of state arise from interactions.  Perturbation theory can be used to study the thermodynamics up to a temperature on the order of $m_s/g_o^2$, where $m_s$ is the string mass scale and $g_o$ is the open string coupling.  Thereafter the perturbative expansion breaks down and higher order terms become important.  In addition, the vacuum energy density is positive and hierachically suppressed with respect to the Planck scale, leading to the possibility that it may contribute to the cosmological constant.

These non-local field theories are known to have vacuum soliton solutions.  In this context soliton refers to a localized nonsingular solution to the classical field equation with finite action.  They have an energy proportional to $m_s/g_o^2$ and therefore will contribute substantially to the partition function for temperatures of that order and higher.  In this paper we study finite temperature classical solutions to the $p$-adic string theories.  These solutions are non-analytic in the open string coupling $g_o$ and therefore cannot be calculated using perturbation theory.  Hence they extend the results obtained in \cite{us1,us2}.

\begin{figure}[!htbp]
\begin{center}
\includegraphics[width=0.54\textwidth,angle=0]{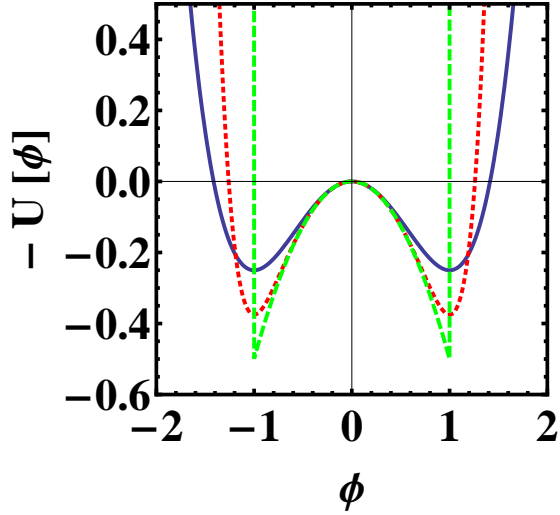}
\end{center}
\caption{(Color online) Inverted potential of the $p$-adic tachyon for $p$ = 3, 7 and $p \rightarrow \infty$. 
\label{potfig}}
\end{figure}

The action for the $p$-adic theory is given by \cite{olson}-\cite{frampton}
\begin{eqnarray}
S &=&
\frac{m_s^D}{g_p^2} \int d^D x  \left[ -\frac{1}{2} \phi\,
{\rm e}^{-{\Box/ M^2}} \phi+\frac{1}{p+1} \phi^{p+1} \right]\,,
\label{action}
\end{eqnarray}
where $\Box = -\partial_t^2 + \nabla_{D-1}^2$ in flat space, and we have defined
\be
\frac{1}{g_p^2} \equiv \frac{1}{g_o^2}\frac{p^2}{p-1}
\;\;\;\; {\rm and} \;\;\;\; M^2 \equiv \frac{2m_s^2}{\ln p} \, .
\label{gpmp}
\ee
The dimensionless scalar field $\phi(x)$ describes the open string tachyon, $m_s$ is the string mass scale, defined by $m_s^2 = 1/2\alpha^{\prime}$ with $\alpha^{\prime}$ the string tension, and $g_o$ is the open string coupling constant.  Though the action (\ref{action}) was originally derived for $p$ a prime number, in this paper we allow it to be any odd integer equal to or greater than 3.  For constant fields, the resulting potential takes the form
\be
U = \frac{m_s^D}{g_p^2} \left(\frac12 \phi^2 - \frac{1}{p+1}\phi^{p+1}\right) \, .
\ee
Its shape is shown in Figure \ref{potfig}.

At finite temperature the scalar field must be a periodic function of imaginary time $\tau$ with period equal to the inverse temperature $\beta = 1/T$, namely, $\phi(\tau + \beta,{\bf x}) = \phi(\tau,{\bf x})$ \cite{KapGale}.  The classical equation of motion is
\be
{\rm e}^{-{\Box/M^2}} \phi = \phi^p
\ee
where $\Box = \partial^2/\partial \tau^2 + \nabla^2$.  Substitution into the action gives $S \equiv -\tilde{S}$ with
\be
\tilde{S} = \frac{1}{2}\left(\frac{p-1}{p+1}\right) \frac{m_s^D}{g_p^2}
\int_V d^{D-1}x  \int_{-\beta/2}^{\beta/2} d\tau
\phi^{p+1}(\tau, {\bf x}) \, .
\ee
The quantity $\tilde{S}$ is obviously non-negative for odd powers $p$.  The classical equation of motion extremizes the action, which in this case means that it minimizes $\tilde{S}$.

All of the finite temperature calculations in our previous papers were done around the perturbative vacuum $\phi=0$, which is of course a trivial solution to the classical equations of motion. Generally, a nontrivial solution $\phi_s$ will have an action with $\tilde{S}_s > 0$.  If, for example, this solution is thought of as a soliton, and the centers of the solitons are widely separated in space-time, then they contribute to the partition function as a dilute gas.  In the usual way \cite{KapGale} this leads to
\be
Z_s = \sum_{n=0}^{\infty} \frac{1}{n!} \left[ \beta V K
\exp(-\tilde{S}_s) \right]^n
\ee
or
\be
\ln Z_s = \beta V K \exp(-\tilde{S}_s) \, .
\ee
Here $K$ is the factor due to quantum fluctuations around the classical solution. The factor $\beta V$ arises because the soliton could be centered anywhere within the spatial volume $V$ and imaginary time interval $\beta$. The nontrivial solutions are exponentially suppressed according to the actual value of their action relative to the trivial solution.  Calculation of the factor $K$ is beyond the scope of this paper.

Since the equation of motion is nonlinear, there may be multiple nontrivial solutions at any given temperature, or none.  Our goal is to find the solutions with the minimum value of $\tilde{S}$ at finite temperature since the others will be exponentially suppressed in comparison.

The outline of our paper is as follows.  In Sect. 2 we derive differential and integral equations to be solved at finite temperature.  In Sect. 3 we extend the diffusion equation method to finite temperature.  In Sect. 4 we deduce from general considerations three types of solution at finite temperature.  In Sects. 5 and 6 we calculate solutions at all temperatures which are even and odd in imaginary time, respectively.  In Sect. 7 we calculate the value of the action for all these solutions to determine which are most important as a function of the temperature.  In Sect. 8 we conclude with a summary of results obtained and potential future directions. 

\section{Differential and Integral Equations}

The classical equation of motion has soliton solutions, as first discovered in Euclidean space at zero temperature \cite{brekke}.  Recall that
\be
{\rm e}^{-{\Box/M^2}} \phi = \phi^p \, .
\ee
This has the trivial solutions $\phi = 0$, $\phi = 1$, and $\phi = -1$ (if $p$ is an odd integer).  Now make the ansatz
\be
\phi(\tau,{\bf x}) = f_0(\tau) f_1(x) f_2(y) f_3(z)
\ee
so that each $f$ satisfies
\be
\exp\left(-\frac{1}{M^2}\frac{d^2}{dx^2}\right)\,f(x) = f^p(x) \, .
\ee
It is easily verified that $f$ must satisfy
\be
f(x) = \frac{M}{2\sqrt{\pi}} \int_{-\infty}^{\infty}
dx' \, {\rm e}^{-M^2 (x'-x)^2/4} f^p(x') \, .
\label{integral-eqn}
\ee
Once again, trivial solutions include $f = 0$, $f = 1$, and $f = -1$ (if $p$ is an odd integer).  A nontrivial solution is
\be
f(x) = \pm p^{1/2(p-1)} {\rm e}^{-(p-1)M^2 x^2/4p} \, .
\label{vacuum soliton}
\ee
Multiplying the $f$'s together results in a soliton in one, two, three, or four dimensions.

For $T > 0$ the scalar field must be periodic in $\tau$ with period $\beta$ \cite{KapGale}.  The ansatz above is applicable, and the solutions in the three space directions are still valid.  In the imaginary time direction the differential equation is still
\be
\exp\left(-\frac{1}{M^2}\frac{d^2}{d\tau^2}\right) f(\tau) = f^p(\tau) \, ,
\label{differential-eqn}
\ee
but now $f(\tau + \beta) = f(\tau)$.

Any periodic function can be expanded in a Fourier series.
\be
f(\tau) = \sum_{n=-\infty}^{\infty} c_n {\rm e}^{i\omega_n \tau}
\ee
Here $\omega_n = 2\pi T n$ is the Matsubara frequency.  Substitution into the differential equation results in
\be
f^p(\tau) = \sum_{n=-\infty}^{\infty} c_n {\rm e}^{\omega_n^2/M^2}
{\rm e}^{i\omega_n \tau} \, .
\ee
On the other hand
\be
\int_{-\beta/2}^{\beta/2} d\tau f^p(\tau) {\rm e}^{-i\omega_m \tau}
= \beta c_m {\rm e}^{\omega_m^2/M^2}
\ee
which implies that
\be
c_n = {\rm e}^{-\omega_n^2/M^2} T \int_{-\beta/2}^{\beta/2} d\tau f^p(\tau)
{\rm e}^{-i\omega_n \tau} \, .
\ee
The integral equation at finite temperature is therefore
\be
f(\tau) = \sum_{n=-\infty}^{\infty} {\rm e}^{-\omega_n^2/M^2}
{\rm e}^{i\omega_n \tau}
T \int_{-\beta/2}^{\beta/2} d\tau'
{\rm e}^{-i\omega_n \tau'}  f^p(\tau') \, .
\label{finiteT inteq}
\ee

One may also expand $f(\tau)$ in sines and cosines instead of complex exponentials.
\be
f(\tau)= A_0 + \sum_{n=1}^{\infty} \left[ A_n \cos(\omega_n \tau)
+ B_n \sin(\omega_n \tau) \right]
\label{trigs}
\ee
In the usual way this leads to
\bd
f(\tau) = T \int_{-\beta/2}^{\beta/2} d\tau' f^p(\tau')
+ 2T \sum_{n=1}^{\infty} {\rm e}^{-\omega_n^2/M^2} \times
\ed
\be
\left[ \cos(\omega_n \tau) \int_{-\beta/2}^{\beta/2} d\tau' f^p(\tau')
\cos(\omega_n \tau') + \sin(\omega_n \tau) \int_{-\beta/2}^{\beta/2}
d\tau' f^p(\tau') \sin(\omega_n \tau') \right] \, .
\label{trig}
\ee
This naturally separates into one integral equation if the solution is even
\be
f_{\rm e}(\tau) = T \sum_{n=-\infty}^{\infty} {\rm e}^{-\omega_n^2/M^2}
\cos(\omega_n \tau) \int_{-\beta/2}^{\beta/2} d\tau' f_{\rm e}^p(\tau')
\cos(\omega_n \tau') \, ,
\label{cos}
\ee
and another if the solution is odd
\be
f_{\rm o}(\tau) = 2T \sum_{n=1}^{\infty} {\rm e}^{-\omega_n^2/M^2}
\sin(\omega_n \tau) \int_{-\beta/2}^{\beta/2} d\tau' f_{\rm o}^p(\tau')
\sin(\omega_n \tau') \, .
\label{sin}
\ee
This separation also follows from eq. (\ref{finiteT inteq}).

The kernel of the integral equation (\ref{finiteT inteq}) is just the theta function of the third kind
\be
\theta_3(u,q)=\sum_{n=-\infty}^{\infty} q^{n^2} {\rm e}^{2uni} \, ,
\ee
so that in more compact notation
\be
f(\tau) = T \int_{-\beta/2}^{\beta/2} d\tau'
\theta_3\left( \pi T (\tau-\tau'), {\rm e}^{-(2\pi T)^2/M^2} \right)
 f^p(\tau') \, .
\ee
Now it is helpful to use the identity
\be
\theta_3\left(u,{\rm e}^{-x^2}\right)
= {\rm e}^{-u^2/x^2} \frac{\sqrt{\pi}}{x}
\theta_3 \left( \frac{i \pi u}{x^2}, {\rm e}^{-\pi^2/x^2} \right)
\label{theta-identity}
\ee
to write
\be
f(\tau) = \frac{M}{2\sqrt{\pi}} \sum_{n=-\infty}^{\infty}
\int_{-\beta/2}^{\beta/2} d\tau' f^p(\tau')
{\rm e}^{-M^2(\tau' - \tau + \beta n)^2/4} \, .
\label{integral-periodic}
\ee
This is an alternative to an expansion in terms of trigonometric functions.

The expression (\ref{integral-periodic}) can be written as
\be
f(\tau) = \frac{M}{2\sqrt{\pi}}\sum_{n=-\infty}^{\infty}
\int_{(n-1/2)\beta}^{(n+1/2)\beta} d\tau_n
f^p(\tau_n-n\beta){\rm e}^{-M^2(\tau_n - \tau)^2/4}
\ee
where $\tau_n\equiv \tau'+n\beta$, then as
\be
f(\tau) = \frac{M}{2\sqrt{\pi}}\sum_{n=-\infty}^{\infty} \int_{(n-1/2)\beta}^{(n+1/2)\beta} d\tau_n
f^p(\tau_n){\rm e}^{-M^2(\tau_n - \tau)^2/4}
\ee
since $f(\tau+\beta)=f(\beta)$, and finally as
\ba
f(\tau)&=& \frac{M}{2\sqrt{\pi}}\sum_{n=-\infty}^{\infty}
\int_{(n-1/2)\beta}^{(n+1/2)\beta} d\tau'
f^p(\tau'){\rm e}^{-M^2(\tau' - \tau)^2/4} \nonumber \\
&=& \frac{M}{2\sqrt{\pi}} \int_{-\infty}^{\infty} d\tau'
f^p(\tau'){\rm e}^{-M^2(\tau' - \tau)^2/4} \, .
\ea
Thus we recover the general integral equation (\ref{integral-eqn}), but (\ref{integral-periodic}) is potentially more useful since it embeds the required periodicity of the solution explicitly while (\ref{integral-eqn}) does not. In the above derivation we assumed that the integral and the summation commutes. This is justified since the infinite sum clearly converges, and the individual integrals are finite (as long as $f$ is nonsingular).

In this section we have derived several integral equations which give solutions to the original differential equation.  These equations may involve either trigonometric functions, such as eqs. (\ref{finiteT inteq}), (\ref{trig}), (\ref{cos}) and (\ref{sin}), or Gaussians, such as eqs. (\ref{integral-eqn}) and (\ref{integral-periodic}).  The original differential equation displays a symmetry where $f \rightarrow f^p$, $f^p \rightarrow f$, and $M^2$ changes sign.  This symmetry may be applied to all of the integral equations too.

\section{Diffusion Equation}

Yet another approach is to cast the problem in terms of a diffusion equation \cite{diffusion1,diffusion2}.  This can be accomplished by introducing an extra dimension labeled by the variable $r$.  The original differential equation is written as
\be
\exp\left(-2r_*\frac{d^2}{d\tau^2}\right) f(\tau) = f^p(\tau) \, ,
\ee
where $r_*=1/2M^2$.  Now introduce a new function $F(\tau,r)$ which satisfies the diffusion equation
\be
\frac{\partial^2}{\partial \tau^2} F(\tau,r) =
\gamma \frac{\partial}{\partial r} F(\tau,r)
\ee
where $\gamma$ is an as yet unspecified constant which can be chosen later.  This function must be periodic in $\tau$ and can be expanded in trigonometric functions just as in eq. (\ref{trigs}) except that now $A_n$ and $B_n$ depend upon $r$.  Substitution into the diffusion equation determines those functions to be
\ba
A_0(r) &=& a_0 \nonumber \\
A_n(r) &=& a_n {\rm e}^{-\omega_n^2 r/\gamma} \;\;\; n \ge 1 \nonumber \\
B_n(r) &=& b_n {\rm e}^{-\omega_n^2 r/\gamma} \;\;\; n \ge 1
\ea
where the $a_n$ and $b_n$ are constants.

The relationship to the original ordinary differential equation is found by examining the operation
\bd
\exp\left(-2r_*\frac{\partial^2}{\partial\tau^2}\right) F(\tau,r) =
\sum_{k=0}^{\infty} \frac{(-2r_*)^k}{k!}
\left(\frac{\partial^2}{\partial \tau^2}\right)^k F(\tau,r)
\ed
\be
= \sum_{k=0}^{\infty} \frac{(-2r_*\gamma)^k}{k!} \left(\frac{\partial}{\partial r} \right)^k F(\tau,r) = F(\tau,r-2r_*\gamma)
\ee
on account of the fact that $F$ obeys the diffusion equation.  Hence the operation of the exponetial operator in $\tau$ is just a translation in the coordinate $r$.  The desired solution $f$ is related to $F$ only at the point $r_*$, namely $f(\tau) = F(\tau,r_*)$.  Hence the proper equation of motion can be phrased as
\be
F(\tau,(1-2\gamma)r_*) = F^p(\tau,r_*) \, .
\ee
Solutions to the original differential equation have been found by evolving some initial nontrivial configuration $F(\tau,0)$ to the final one $F(\tau,r_*)$ via the diffusion equation.  Although this is an interesting approach, so far we have not found any advantage over solution of the original differential or integral equations in one variable.

\section{General Considerations}

One can prove some general theorems analogous to the theorems for periodic solutions in real time \cite{rolling}.  Suppose we have a periodic solution satisfying
\be
f_{\rm min} \leq f(\tau) \leq f_{\rm max}
\ee
with $f_{\rm min} < f_{\rm max}$.  Let us first assume that $f_{\rm max} > 0$.  Then using eq. (\ref{integral-eqn}) we find that
\be
f(\tau) <  \frac{M}{2\sqrt{\pi}} \int_{-\infty}^{\infty}
d\tau' \, {\rm e}^{-M^2 (\tau'-\tau)^2/4} f_{\rm max}^p =f_{\rm max}^p \, .
\ee
Taking the maximum value of $f(\tau)$ on the left side we infer that
$f_{\rm max} < f_{\rm max}^p$, which further implies that $f_{\rm max} > 1$.  On the other hand, if $f_{\rm max} < 0$ then we infer that $-1 < f_{\rm max} < 0$.

Now consider the lower limit of the oscillation.
\be
f(\tau) >  \frac{M}{2\sqrt{\pi}} \int_{-\infty}^{\infty}
d\tau' \, {\rm e}^{-M^2 (\tau'-\tau)^2/4} f_{\rm min}^p = f_{\rm min}^p\, .
\ee
First assume that $f_{\rm min} > 0$.  Taking the minimum value of $f(\tau)$ on the left side we infer that $f_{\rm min} > f_{\rm min}^p$, which further implies that $f_{\rm min} < 1$.  On the other hand, if $f_{\rm min} < 0$ then we infer that $f_{\rm min} < -1$.

Therefore, we conclude that there could exist three types of oscillations.  (i) $0 < f_{\rm min} < 1$ and $f_{\rm max} > 1$.  (ii) $f_{\rm min} < -1$ and $f_{\rm max} > 1$.  (iii) $f_{\rm min} < -1$ and $-1 < f_{\rm max} < 0$.  These results are rather intuitive, and correspond to oscillations about $\phi = 1$, $\phi = 0$, and $\phi = -1$, respectively.  Furthermore, note that there are conditions on the amplitudes of the oscillations, which are a result of the nonlinearity of the original differential equation.

\section{Even Solutions in Imaginary Time}

In this section we consider solutions that are even in $\tau$.  First consider the low temperature limit when $T \ll M$.  Then the width of a vacuum soliton, $1/M$, is much less than $\beta$.  This means that one can have a dilute gas of solitons.  When $-\beta/2 < \tau < \beta/2$, only the $n=0$ term will contribute significantly in eq. (\ref{integral-periodic}).  When $\tau \approx \beta$, only the $n=1$ term will contribute.  When $\tau \approx 2\beta$, only the $n=2$ term will contribute.  And so on and so forth.  Then the periodic approximate solution to the integral and differential equations is
\be
f(\tau) = \pm p^{1/2(p-1)} \sum_{n=-\infty}^{\infty}
{\rm e}^{-(p-1)M^2 (\tau - \beta n)^2/4p} \, .
\label{lowtemp-soli}
\ee
In the limit $\beta \rightarrow \infty$ with $-\beta/2 \ll \tau \ll \beta/2$ only the $n=0$ term survives and the vacuum solution (\ref{vacuum soliton}) is recovered.

An alternative approach in the low temperature limit is to substitute the vacuum solution into eq. (\ref{cos}) and extend the limits of integration to infinity.  This gives
\ba
f(\tau) &=& \pm 2 \, p^{1/2(p-1)} \sqrt{\frac{p\pi}{p-1}} \frac{T}{M}
\sum_{n=-\infty}^{\infty} {\rm e}^{-p\omega_n^2/(p-1)M^2}
\cos(\omega_n \tau) \nonumber \\
&=& \pm 2 \, p^{1/2(p-1)} \sqrt{\frac{p\pi}{p-1}} \frac{T}{M}
\theta_3\left( \pi T \tau, {\rm e}^{-p(2\pi T)^2/(p-1)M^2} \right)
 \, .
\ea
Applying the identity (\ref{theta-identity}) to this results in eq. (\ref{lowtemp-soli}).

Since we have seen that oscillations can only exist around the points $\phi = 0, \pm 1$, let us start by looking at small oscillations around $\phi=1$. To capture these let us make a truncated harmonic series expansion.
\be
f(\tau)=A_0 + A_1 \cos(\omega \tau)
\label{harmonic}
\ee
Substituting (\ref{harmonic}) in the field equation (\ref{differential-eqn}), and using trigonometric identities, we find that at the same level of truncation
\ba
A_0^{p-1} + \frac{p(p-1)}{4} A_0^{p-3} A_1^2 &=& 1 \\
p A_0^{p-1} + \frac{p(p-1)(p-2)}{8} A_0^{p-3} A_1^2&=&
{\rm e}^{\omega^2/M^2} \, .
\ea
In the lowest order approximation, $A_1^2 \ll 1$, we find the usual harmonic oscillations around the minimum to be
\ba
A_0 &=& 1 \nonumber \\
\omega^2 &=& M^2\ln p \, .
\ea
Because the solutions must be periodic in imaginary time with period $\beta$ we must have $\omega = 2\pi m T$ where $m$ is an integer which, without loss of generality, we may assume to be positive.  The smallest frequency corresponds to $m=1$.  This means that the solitons correspond to only a particular temperature $T_c \equiv M\sqrt{\ln p}/2\pi$. How do we get solitons with other temperatures? This is where the anharmonicity comes into play and one realizes that the frequency changes as we change the amplitude of the oscillations. Going to the next order we have
\ba
A_0 &=& 1-\frac{p}{4} A_1^2 + \cdot\cdot\cdot \\
\omega^2 &=& M^2 \ln \left( p-\frac{p(p+2)(p-1)}{8} A_1^2
+ \cdot\cdot\cdot \right) \, .
\ea
This means that the temperature decreases as the amplitude is increased.

In the above analysis higher harmonics were neglected.  The first overtone can be included by writing
\be
f(\tau)=A_0 + A_1 \cos(\omega \tau) + A_2 \cos(2\omega \tau)
\label{overtone}
\ee
Proceeding as before we find
\be
f(\tau) = 1 - \oneqt p A_1^2 + A_1 \cos(\omega \tau)
+ \frac{1}{4(p^2 + p + 1)} A_1^2 \cos(2\omega \tau)
+ \cdot\cdot\cdot \, .
\ee
The expression for the frequency in terms of the amplitude is the same as before.

The solution that is even in $\tau$ evolves with temperature in the following way.  At $T=0$ the soliton solution is the Gaussian given by eq. (\ref{vacuum soliton}).  When $T \ll M$ a periodic solution is constructed from widely spaced Gaussians as expressed by eq. (\ref{lowtemp-soli}).  As the temperature increases further the tails of the Gaussians begin to overlap and the solution becomes more uniform.  Eventually it evolves into the form of eq. (\ref{overtone}) with $A_0 \gg A_1 \gg A_2$ and $A_0 \rightarrow 1$.  The solution goes to the constant 1 at the well-defined critical temperature $T_c=M\sqrt{\ln p}/2\pi$; thereafter there is only the trivial solution $f=1$.  Exactly the same evolution happens for negative $f$.  This behavior is in accord with the general considerations delineated earlier.  

Precise solutions for the full range of $T$ can only be done numerically.  
It is convenient to use dimensionless variables $t \equiv T/M$ and $u \equiv T \tau$.  The integral equation to solve, eq. (\ref{cos}), assumes the solution is an even function.
\be
f(u) = 2 \int_0^{1/2} du' f^p(u') + 4
\sum_{n=1}^{\infty} {\rm e}^{-(2\pi n t)^2}
\cos(2\pi n u) \int_0^{1/2} du' f^p(u')
\cos(2\pi n u')
\label{scaled-cos}
\ee
There are no general methods for solving nonlinear differential or integral equations, and the fact that the differential equation under consideration has an infinite number of derivatives means that the problem is infinitely non-local.  A straightforward iteration of the integral equation, starting with a low temperature or high temperature trial solution, does not converge.  The reason is that the solutions always have a region where $f>1$, and if the trial solution is too large by a small amount then raising it to a power $p>1$ will take it even further away from the true solution.  An important observation is to note a constraint that follows from eq. (\ref{scaled-cos}), which is
\be
\int_0^{1/2} du f(u) = \int_0^{1/2} du f^p(u) \, .
\ee
The method used here starts with a trial solution and iterates, where after each iteration, the new solution is scaled by a constant to satisfy this constraint.  In this way convergence is achieved quite rapidly.  Typically we stop the iteration after an accuracy of at least eight significant figures is attained.

\begin{figure}[htbp]
\begin{center}
\includegraphics[width=0.8\textwidth,angle=0]{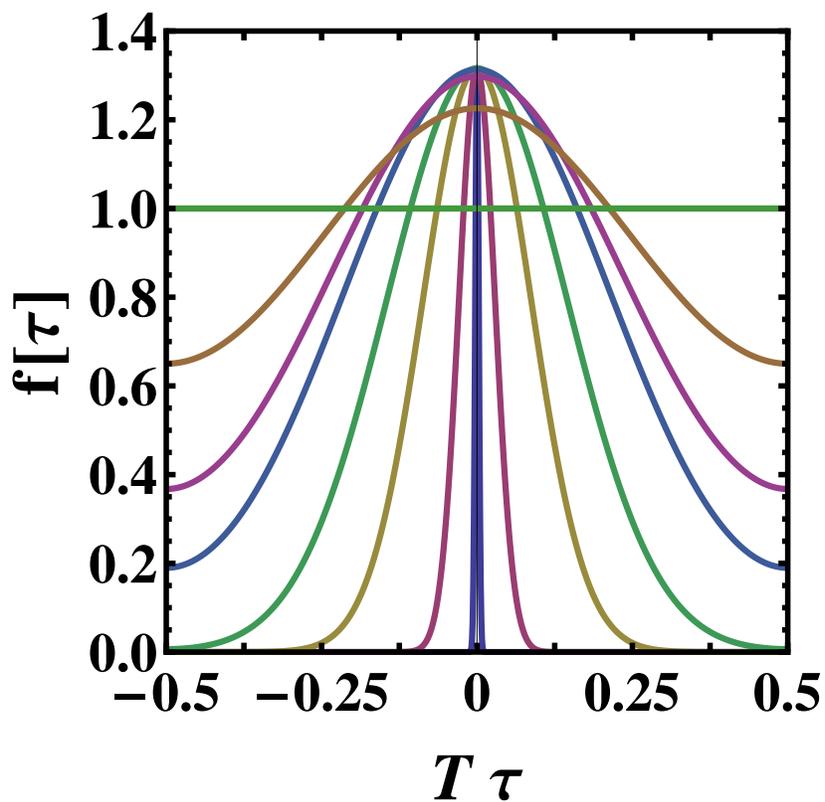}
\end{center}
\caption{(Color online) Even soliton profile for $p$ = 3 as a function of imaginary time $\tau$ for different temperatures: $T/T_c = 0.01, \, 0.10, \, 0.30, \, 0.50, \, 0.75, \, 0.85, \, 0.95$.  For $T/T_c \ge 1$ the solution is exactly 1.  As the temperature increases, the solution transitions from a narrow Gaussian, to a cosine wave, to the constant 1 at $T_c$.
\label{even1}}
\end{figure}

\begin{figure}[htbp]
\begin{center}
\includegraphics[width=0.75\textwidth,angle=0]{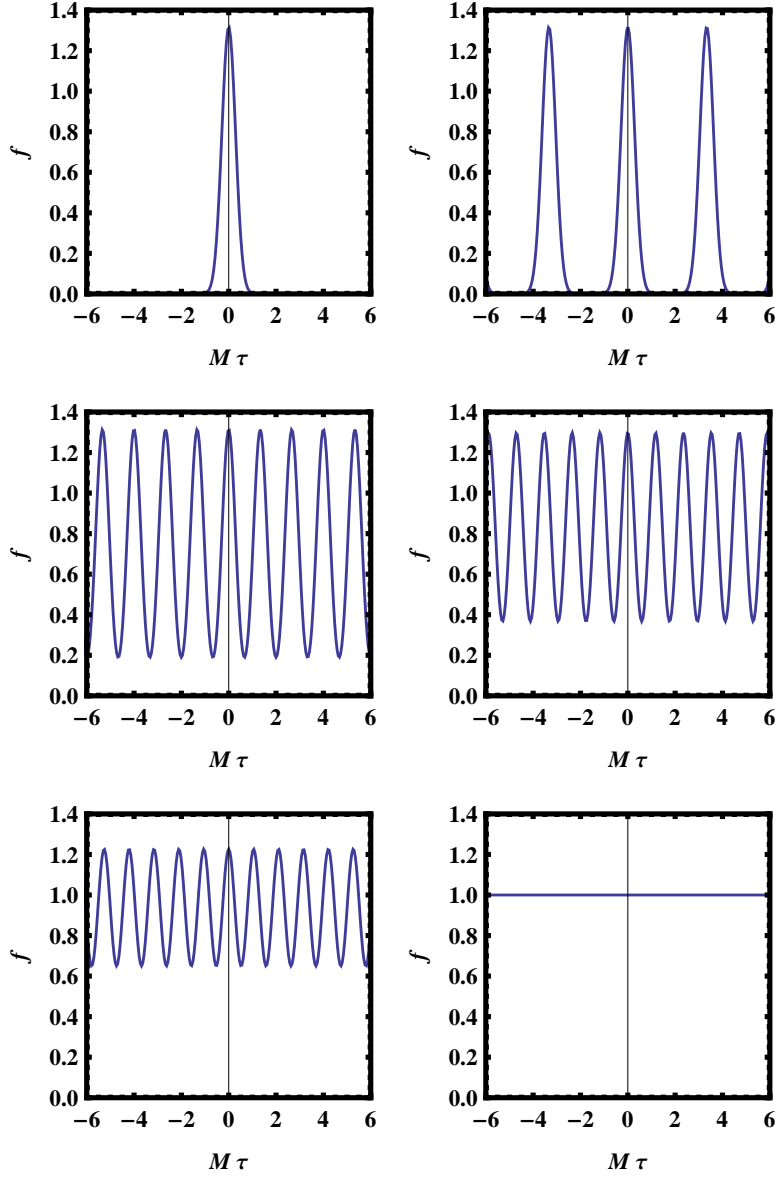}
\end{center}
\caption{(Color online) Even soliton profile for $p$ = 3 as a function of imaginary time $\tau$ for different temperatures: $T/T_c = 0.10, \, 0.30, \, 0.75, \, 0.85, \, 0.95, \, 1.00$, from top left to bottom right, in units of the p-adic string scale $M$.
\label{even2}}
\end{figure}

Some illustrative results are plotted in Fig. \ref{even1} as a function of 
$T \tau$, and in Fig. \ref{even2} as a function of $M \tau$.  Referring to Fig. \ref{even1}, the profile is described very well by the Gaussian for $T < 0.5 T_c$.  It then transitions to a cosine before flattening to the constant 1 for $T \ge T_c$.  This is exactly the behavior predicted by the analytical approximations.

\section{Odd Solutions in Imaginary Time}

Now consider an oscillation that is odd in $\tau$.  Such an oscillation can only occur about $\phi = 0$.  The harmonic expansion starts out as
\be
f(\tau) = B_1 \sin(\omega \tau) + B_3 \sin(3\omega \tau) \, .
\label{highTodd}
\ee
In order that the solution have the required periodicity implies that $\omega = 2\pi (2m+1)T$ where, without loss of generality, $m = 0, 1, 2, ...$.  Substitution into the differential equation, and matching only the
$\sin(\omega \tau)$ term, results in
\be
B_1 = \left[ \frac{(p+1)!!}{2 p!!} \right]^{1/(p-1)}
{\rm e}^{\omega^2/(p-1)M^2} \, .
\ee
Matching both the fundamental and first overtone results in
\ba
B_1 &=& \left[ \frac{(p+1)!!}{2 p!!} \right]^{1/(p-1)}
{\rm e}^{\omega^2/(p-1)M^2} \left\{ 1 - \frac{p(p-1)}{(p+3)^3}
{\rm e}^{-8\omega^2/M^2} \right\}
\nonumber \\
B_3 &=& \left[ \frac{(p+1)!!}{2 p!!} \right]^{1/(p-1)}
{\rm e}^{\omega^2/(p-1)M^2} \left\{ -\left( \frac{p-1}{p+3} \right)
{\rm e}^{-8\omega^2/M^2}  \right\}\, .
\ea
This solution has the characteristic feature predicted previously; namely, the amplitude of oscillation exceeds 1.  In fact, the amplitude grows exponentially as $T^2/M^2$, and the fundamental frequency dominates at large temperature.  Although solutions exist for any integer value of $m$, the solution with $m=0$ is the most important because the others have actions which are exponentially larger.

When $T > M/4\pi$ the first term in the series (\ref{highTodd}) dominates.  As $T$ decreases the amplitude decreases, and more and more terms in the series become important.  There is no known nontrivial solution at $T=0$, so it is interesting to see what happens in the limit $T \rightarrow 0$.  In fact, one can construct a solution both periodic and odd in imaginary time by adding the vacuum Gaussian solutions in the following way.
\be
f(\tau) = p^{1/2(p-1)} \sum_{n=-\infty}^{\infty} (-1)^n
{\rm e}^{-(p-1)M^2 (\tau - \beta (n+1/2))^2/4p} \, .
\label{low-odd}
\ee
When $T < M/4\pi$, these Gaussians are spaced much further apart than their widths, and so the differential and integral equations are satisfied to very good approximation.  In particular, there is a positive Gaussian centered at $\tau = \beta/2$ and a negative Gaussian centered at $\tau = -\beta/2$.  As $T$ increases these naturally go over to $B_1 \sin(2\pi T \tau)$.  As $T$ goes to zero, the spacing in $\tau$ between adjacent Gaussians diverges as $\beta = 1/T$.

Once again, precise solutions at intermediate temperatures require numerical calculation.  For an odd solution the equation to solve is
\be
f(u) = 4 \sum_{n=1}^{\infty} {\rm e}^{-(2\pi n t)^2}
\sin(2\pi n u) \int_0^{1/2} du' f^p(u')
\sin(2\pi n u') \, .
\label{scaled-sin}
\ee
In this case the constraint we use takes the form
\be
\int_0^{1/2} du f(u) \sin(2\pi u) =
{\rm e}^{-(2\pi t)^2} \int_0^{1/2} du f^p(u) \sin(2\pi u) \, .
\ee
As with the even solutions, convergence is achieved rather rapidly.

\begin{figure}[htbp]
\begin{center}
\includegraphics[width=0.8\textwidth,angle=0]{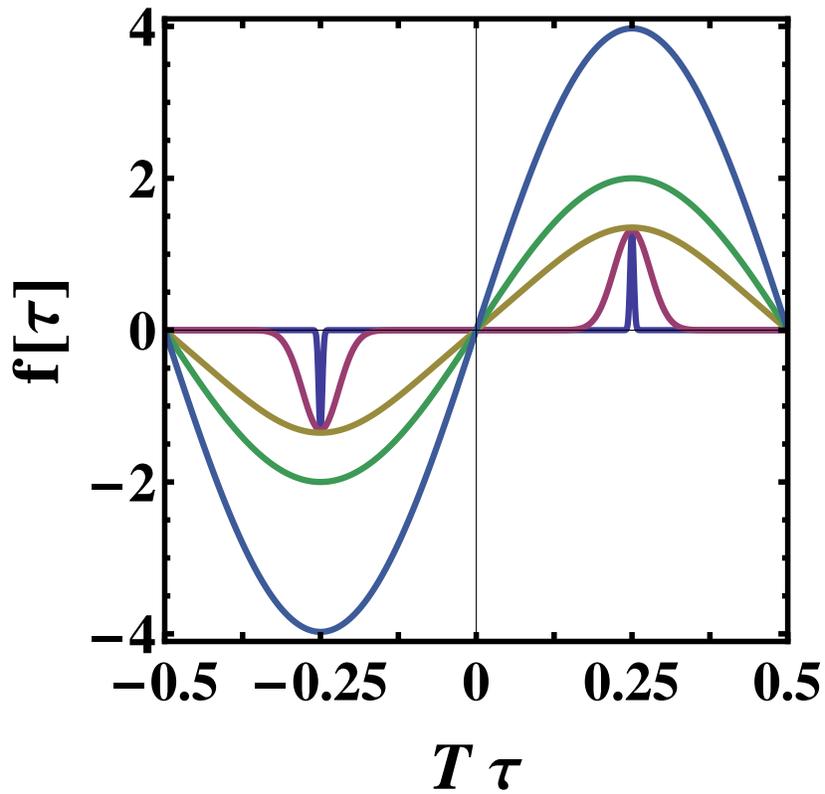}
\end{center}
\caption{(Color online) Odd soliton profile for $p$ = 3 as a function of imaginary time $\tau$ for different temperatures: $T/T_c = 0.01, \, 0.10, \, 0.50, \, 1.00, \, 1.50$.  As the temperature increases, the solution transitions from a pair of narrow Gaussians to a sine wave whose amplitude increases with temperature.
\label{odd1}}
\end{figure}

\begin{figure}[htbp]
\begin{center}
\includegraphics[width=0.75\textwidth,angle=0]{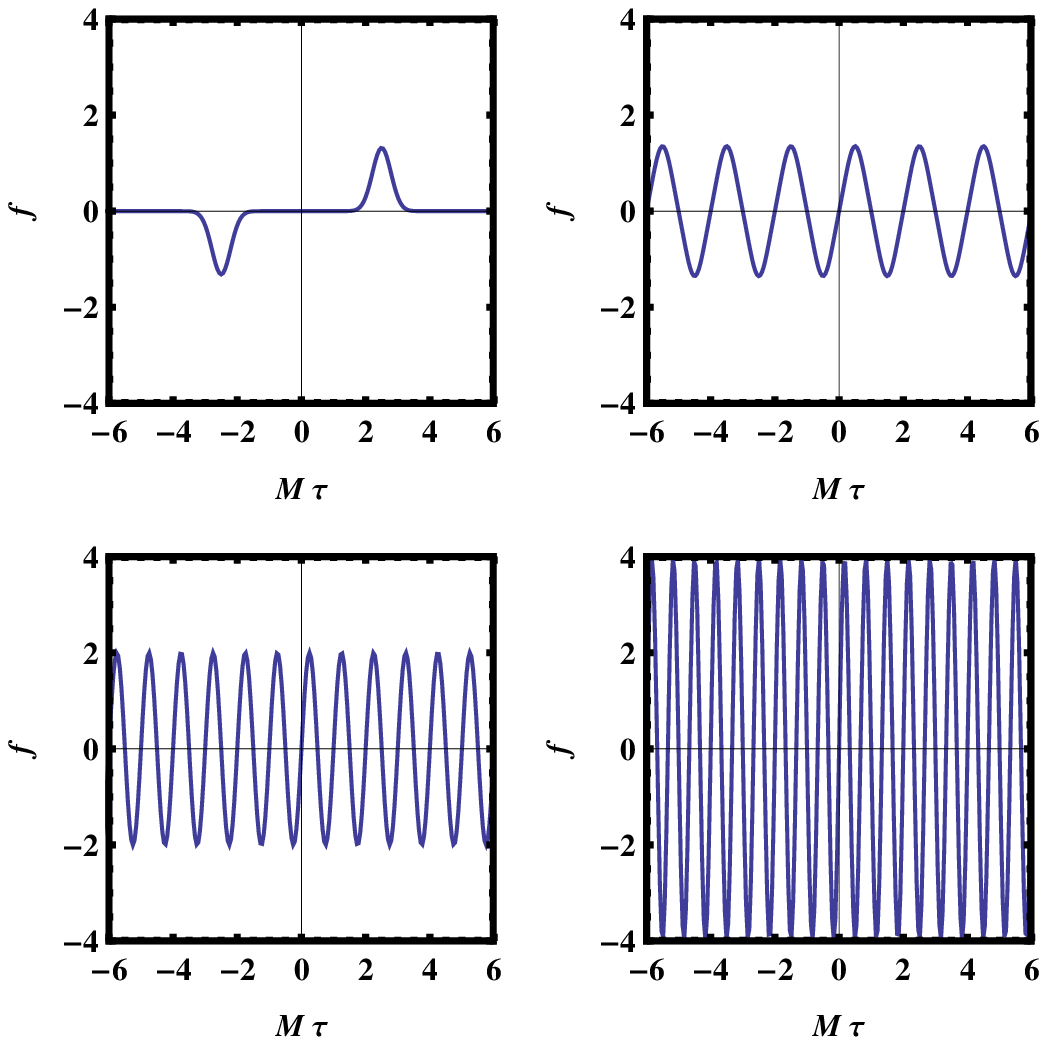}
\end{center}
\caption{(Color online) Odd soliton profile for $p$ = 3 as a function of imaginary time $\tau$ for different temperatures: $T/T_c = 0.10, \, 0.50, \, 1.00, \, 1.50$, from top left to bottom right, in units of the p-adic string scale $M$.
\label{odd2}}
\end{figure}

Some illustrative results are plotted in Fig. \ref{odd1} as a function of 
$T \tau$, and in Fig. \ref{odd2} as a function of $M \tau$.  Referring to Fig. \ref{odd1}, the profile is described very well by a pair of Gaussians for $T < 0.5 T_c$.  It then rapidly transitions to a sine with increasing amplitude for $T > T_c$.  This is exactly the behavior predicted by the analytical approximations.

\section{Action}

As discussed in the introduction, the contribution of solitons to
$\ln Z$ decreases exponentially with their action.  The spatial
distribution of the solitons is always given by the Gaussian of Eq.
(\ref{vacuum soliton}).  Inserting this into the action gives
\be \tilde{S}(T) = \frac{1}{2}\left(\frac{p^2}{p+1}\right)
\frac{1}{g_o^2} \left[2\pi \frac{\ln p}{p^2 -1} p^{2p/(p-1)}
\right]^{(D-1)/2} I(T) \, ,
\ee
where the dimensionless integral
\be 
I(T) \equiv m_s \int_{-\beta/2}^{\beta/2} d\tau
f^{p+1}(\tau) 
\ee
determines the temperature dependence.

First consider the low temperature limit.  For $T < 0.5 T_c$ the even solution gives 
\be
I(T) = \sqrt{ \frac{2\pi \ln p}{p^2-1}} \, p^{p/(p-1)}
\ee
to good approximation. Meanwhile the odd solution gives  
\be
I(T) = 2 \sqrt{ \frac{2\pi \ln p}{p^2-1}} \, p^{p/(p-1)}
\ee
to good approximation, which is twice the value of the even solution.  This is because there are two Gaussian distributions inside the interval of integration. See Figs. \ref{even1} and \ref{odd1}.  

Now consider the high temperature limit.  For $T \ge T_c$ the exact result for the even solution is
\be
I(T) = \frac{\sqrt{2} \pi T_c}{T} \, . 
\ee
For $T > T_c$ the odd solution gives
\be
I(T) = \frac{\pi T_c}{\sqrt{2}T} \left[ \frac{(p+1)!!}{2p!!} \right]^{2/(p-1)}
\exp\left[\left(\frac{p+1}{p-1}\right) \left(\frac{2\pi T}{M}\right)^2 \right] \ee
which increases exponentially in $T^2/M^2$.

We can compute the contributions from the classical soliton configurations numerically. These are displayed in Fig. \ref{action}.  As discussed above, the high temperature behaviors are opposite. The action for the even soliton solution decreases as $1/T$ for $T > T_c$ while the action for the odd soliton solution grows exponentially. At low temperature the action for the odd solution is twice that of the even solution.  From these computations, we can deduce that the odd soliton contribution is unimportant at all temperatures, whereas the even soliton contribution is important when $T > m_s/g_o^2$.

\begin{figure}[!htbp]
\begin{center}
\includegraphics[width=3.0in,angle=0]{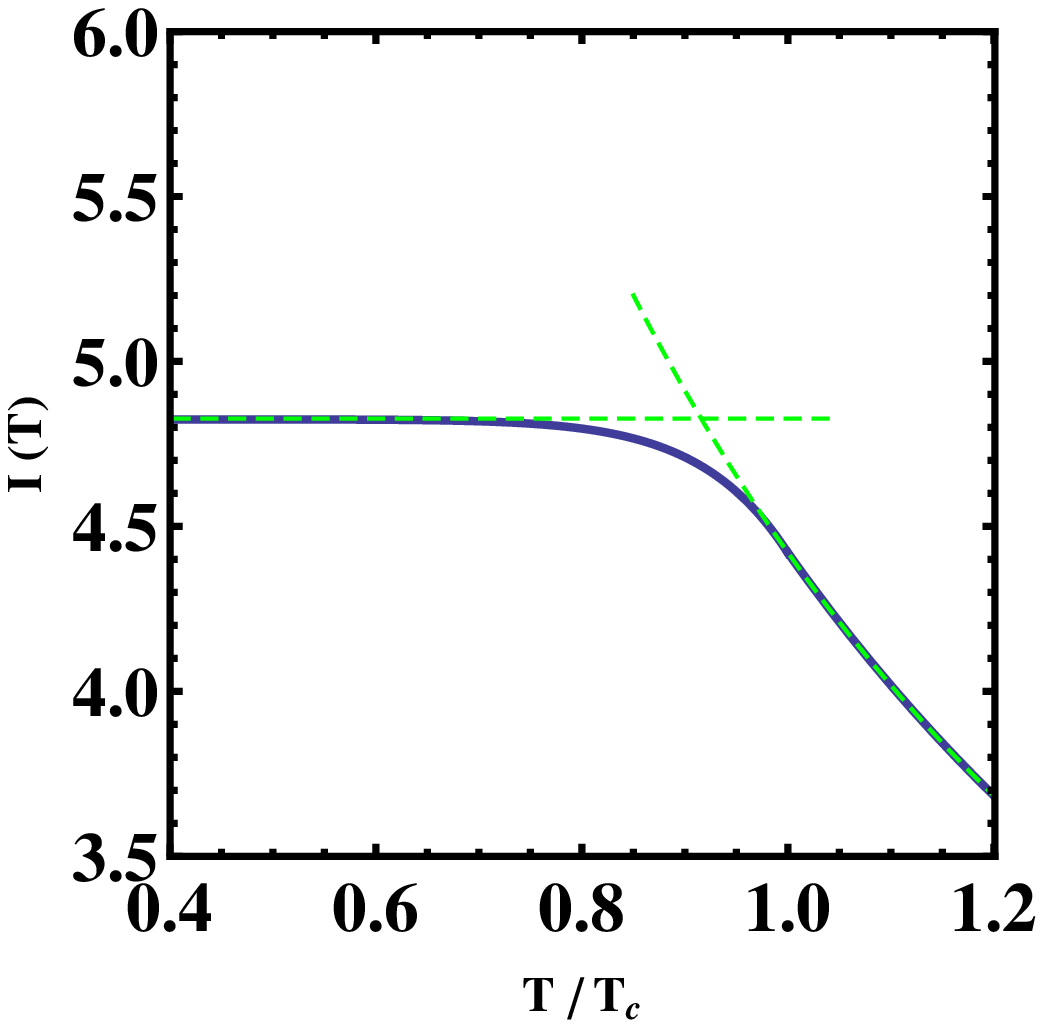}
\end{center}
%\vspace{0.1in}
\begin{center}
\includegraphics[width=3.0in,angle=0]{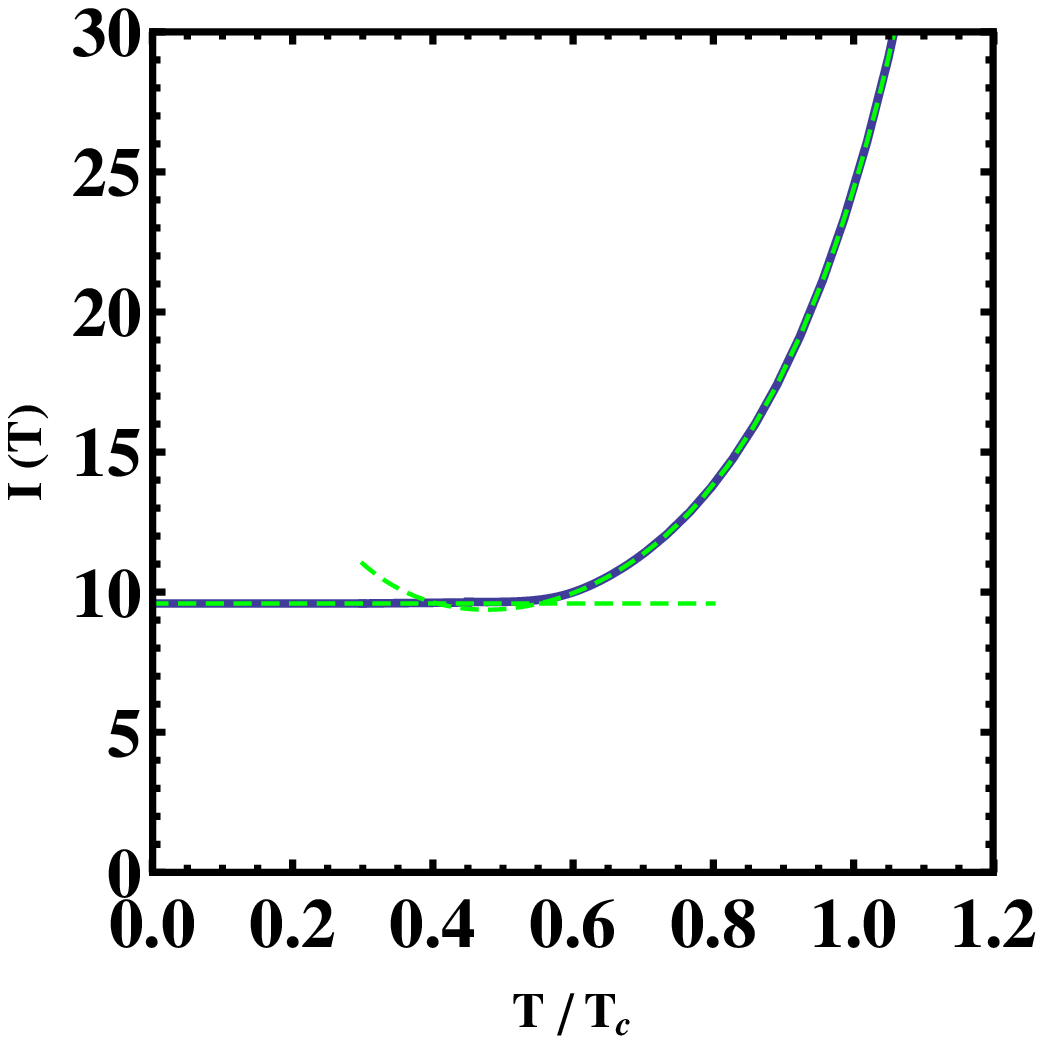}
\end{center}
\caption{(Color online) Contribution $I(T)$ to the Euclidean action of the even (top) and odd (bottom) solitons for $p$ = 3 as a function of $T/T_c$. The green dashed lines show the asymptotic behaviour at low and high temperatures.}
\end{figure}

\newpage

\section{Discussion}

In this paper we have found finite temperature soliton solutions to a class of non-local field theories arising from $p$-adic string theory.  Analytic solutions were derived for both low and high temperatures, delineated by a critical temperature $T_c \sim m_s/g_o^2$.  In the intermediate temperature region numerical solutions were readily found. 

These soliton solutions have finite action even at zero temperature.  This means that they will contribute to the vacuum energy density or cosmological constant.  However, their contribution is suppressed exponentially as 
${\rm e}^{-c_p/g_o^2}$, where $c_p$ is a number dependent upon $p$, relative to the perturbative contribution \cite{us2}.  Hence they will be unimportant for open string couplings $g_o \ll 1$. 

In order to quantify the soliton contributions to the partition function, quantum fluctuations around the classical solutions must be calculated.  That work is in progress.  

\section*{Acknowledgements}

This work was supported by the U.S. DOE Grant Nos. DE-FG02-87ER40328 and DOE/DE-FG02-94ER40823, the FPA 2005-02327 project (DGICYT, Spain), and the CAM/UCM 910309 project.

\end{document}